\documentstyle[prl,aps,floats]{revtex}
\tighten
\def\be{\begin{equation}}

\def\b{\beta}
\def\a{\alpha}
\def\g{\gamma}
\def\s{\sigma}

\def\e{\epsilon}
\def\d{\delta}

\def\bear{\begin{eqnarray}} 
\def\ee{\end{equation}}
\def\eear{\end{eqnarray}}
\def\pa{\partial}

\begin{document}
\twocolumn[\hsize\textwidth\columnwidth\hsize\csname @twocolumnfalse\endcsname
\title{The commutativity principle and lagrangian symmetries}
\author{R. Banerjee\cite{r}}
\address{
S.N. Bose National Centre For Basic Sciences\\
She commutativity principalt Lake City, Block JD, Sector III\\
Calcutta-700 091, INDIA\\
}
\maketitle
\begin{abstract}
Using the commutativity of a general variation with the time differentiation
we discuss both global and local (gauge) symmetries of a lagrangian from a
unified point of view. The Noether considerations are thereby applicable
for both cases.
A complete equivalence between the hamiltonian and lagrangian
formulations is established.
\end{abstract} 
\vskip 0.5cm
]
An important problem is the study of various symmetries  
of a given action. Thus, for example, global symmetries, 
or gauge invariances of the first kind are crucial for condensed 
matter systems, whereas local symmetries, or gauge 
invariances of the second kind pervade the whole of gauge theories.
Symmetry transformations are those transformations 
that keep the invariance of the action without using the 
equations of motion.The quantum mechanical implementation of 
these symmetry principles is naturally
carried out in the lagrangian formalism since the equations of motion 
assume the form of a variational principle. This is how the global
symmetries are usually studied, leading to conservation
laws using Noether's theorem.

Local (gauge) symmetries,  
on the other hand, are best understood in the hamiltonian
formalism by using the procedure of Dirac \cite{Dirac} to identify the constraints,
which are a consequence of the gauge freedom of the
theory. The generator is constructed as a linear combination of these 
constraints. For this generator to act as a symmetry of the action, there have
to be certain conditions \cite{{Henn},{Ban1},{Ban2}} among the parameters entering in the definition of
the generator. Alternatively, there exist purely  lagrangian methods \cite{{Git},{Cha},{Sh}} of 
extracting the gauge symmetry, but the connection with the hamiltonian approach
remains obscure, just as the meaning of Noether's theorem in this context remains unclear.

In this paper we present a unified approach to the implementation of either
global or local symmetries. It is based on the commutativity of a general
variation with the time differentiation operation, which was
used by us \cite{Ban1} recently to discuss certain aspects of local symmetries.
An analogue of Noether's theorem is obtained.
A complete equivalence between the lagrangian and hamiltonian
formalisms is shown. 
With this in mind we will consider first order systems since here both the
lagrangian and hamiltonian formalisms can be applied rightaway. This is not 
a serious restriction since any second order lagrangian can always be brought
to a first order form by a suitable extension of the configuration space.

Consider the following lagrangian,
\be
L= a^\a(q)\dot q_\a - V(q)
\label{lag}
\ee
where the first and second terms denote the kinetic and potential pieces,
respectively.
Note that the $a_\a(q)(\a=1,...N)$ includes the extra variables that might be needed to
recast the lagrangian into a first order form. Under an infinitesimal 
variation $\d q_\a$, the lagrangian changes as,
\be
\d L= \Big(f_{\a\b}\dot q^\b - \frac{\pa V}{\pa q^\a}\Big)\d q^\a
\label{var}
\ee
where the symplectic two form is given by,
\be
f_{\a\b}=\frac{\pa a_\b}{\pa q^\a}-\frac{\pa a_\a}{\pa q^\b}
\label{symplectic}
\ee
If the above matrix is invertible, then the equations of motion for all the
coordinates can be determined in the usual way from the invariance of the
action,
\be
\dot q^\a = f^{\a\b}\frac{\pa V}{\pa q^\b}
\label{motion}
\ee
where $f^{\a\b}$ is the inverse of $f_{\a\b}$ defined as $f_{\a\b}f^{\b\g}=\delta^\g_\a$.

If the matrix (\ref{symplectic}) is not invertible, then the equations of
motion cannot be determined for all the coordinates. In other words the 
number of equations of motion is less than the number of variables so that
there is a degeneracy in the Cauchy problem. This corresponds to the case
of a gauge theory where local symmetries play an important role. The global
symmetries can be discussed without this additional complication and so we
turn to the case where the symplectic matrix is invertible.

Just to show the equivalence between the lagrangian and 
hamiltonian formalisms, recall that the basic brackets are given by the
inverse of the symplectic matrix,
\be
\{a^\a, a^\b\}=\frac{\pa a^\a}{\pa q^\g}f^{\g\s}\frac{\pa a^\b}{\pa q^\s}
\label{bracket}
\ee
The equation of
motion (\ref{motion}) is now expressed as,
\be
\dot q_\a=\{q_\a, V\}
\label{hameqn}
\ee
thereby yielding the conventional hamiltonian form of the equation of motion.

An important step in obtaining (\ref{var}) from (\ref{lag}) has been to use the
commutativity of the general variation with the time derivative operation,
\be
\delta\Big(\frac{dq_\a}{dt}\Big)=\frac{d}{dt}\Big(\delta q_\a\Big)
\label{comm}
\ee
This relation is crucial for deriving the lagrange's equation. In the hamiltonian 
context, it leads to nontrivial consequences. Under a global transformation,
the variation $\delta q_\a$ is given by,
\be
\d q_\a=\{q_\a, G\}
\label{generator}
\ee
Independently computing both sides of (\ref{comm}) by using (\ref{hameqn})
and (\ref{generator}), and then exploiting the Jacobi identity yields,
$$
\{q_\a, \{G, V\}\}=0
$$
Since the result is true for all $q_\a$, we can make the stronger statement that,
\be
\{G, V\}=0
\label{law1}
\ee
It is easy to see that this 
condition leads to the off-shell invariance of the action, since using
(\ref{generator}) in (\ref{var}), the expression for $\delta L$ reduces, modulo
a total time derivative, to $\{G, V\}$. The conservation laws following from the
global symmetries require the explicit use of the equations of motion. Hence (\ref{law1})
may be reexpressed as,
\be
\frac{dG}{dt} =0
\label{law}
\ee
thereby reproducing the usual statement of Noether's theorem regarding the conservation
of the generator.

Let us next discuss the case of local symmetries. 
As stated before the symplectic matrix is not invertible and there are
constraints in the system related to these noninvertible velocities.
As shown by Dirac, the action principle for a constrained system follows
from a lagrangian with a general structure,
\be
L=a^i(q)\dot q_i-\lambda^a\phi_a(q)-V(q)
\label{local}
\ee
The coordinates $q_i (i=1,...n)$ constitute the nonsingular part of the 
original lagrangian (\ref{lag}) while the constraints $\phi_a (a=1,...N-n)$ are implemented by the
lagrange multipliers $\lambda_a$.
For standard (i.e. second order) lagrangians the momenta 
corresponding to the noninvertible velocities are defined to tbe the 
primary constraints of the theory. The other (secondary, tertiary
, etc) constraints are
obtained from the successive time consistency of 
these constraints till the iterative process terminates. This is the 
Dirac algorithm in the hamiltonian approach. In passing to the first order
form the primary constraints are naturally eliminated and the variables (say $\lambda{a_1}$)
conjugate to these constraints impose the secondary
constraints ($\phi_{a_1}$).
All the other constraints (labelled by the index ($a_2$)) are put in by hand 
through their corresponding (unknown) lagrange multipliers $\lambda_{a_2}$. The complete
set of constraints is then labelled by the index $a$, which is a sum of $a_1$
and $a_2$.
 
Under an arbitrary variation, the lagrangian (\ref{local}) changes as,
$$
\d L=\Big( f_{ji}\dot q^i-\lambda^a\frac{\pa\phi_a}{\pa q^j}-\frac
{\pa V}{\pa q^j}\Big)\d q^j-\phi_a\d\lambda^a
$$
The symplectic matrix is now invertible and the invariance of the action
leads to the following equations of motion,
$$
\dot q^i=f^{ij}\Big(\lambda^a\frac{\pa\phi_a}{\pa q^j}+\frac
{\pa V}{\pa q^j}\Big)
$$
\be
\phi_a=0
\label{constraint}
\ee
which can also be put in the hamiltonian form by using the brackets,
\be
\dot q_i=\{q_i, \phi_a\}\lambda^a + \{q_i, V\}
\label{neweqn}
\ee
Since the full set of constraints has been found, consistency with the
equations of motion demands that they satisfy the algebra,
\bear
\{\phi_a, \phi_b\}&=&C_{ab}^c \phi_c\nonumber \\
\{V, \phi_a\}&=&V_a^b\phi_b\label{algebra}\eear

The structure of the above algebra shows that the constraints act as the 
generator in the hamiltonian framework in the sense that the hamiltonian has
vanishing brackets with these constraints. Indeed following Dirac, the usual
form of this generator is taken as,
\be
G=\epsilon^a\phi_a
\label{hamgen}
\ee
where $\e^a$ are the corresponding parameters and the infinitesimal transformations
are given by,
\be
\d q_i=\{q_i, \phi_a\}\e^a
\label{transfm}
\ee
We are now in a position to discuss the local gauge invariance of (\ref{local}).
Computing the l.h.s. of (\ref{comm}) from (\ref{neweqn}) by using 
(\ref{transfm}), we obtain,
\be
\d \dot q_i=\{\{q_i, \phi_a\},\phi_b\}\e^b\lambda^a+
\{\{q_i, V\},\phi_b\}\e^b+\{q_i, \phi_a\}\d\lambda^a
\label{delta}
\ee
Note that the expression $\d\lambda_a$ is only formal since in general we do not
know the lagrange multipliers. Its precise meaning will be abstracted
from the commutativity law (\ref{comm}).
Next, the r.h.s. of (\ref{comm}) is computed independently
and equated with (\ref{delta}).
Exploiting the Jacobi identity we find,
$$
\{q_i, \phi_a\}\Big(\d\lambda^a-\dot\e^a+C^a_{cb}\lambda^c\e^b+V^a_b\e^b\Big)=0
$$
Since the result is valid for all $q_i$, we get the following transformation law
for the multipliers,
\be
\d\lambda^a-\dot\e^a+C^a_{cb}\lambda^c\e^b+V^a_b\e^b=0
\label{dellambda1}
\ee
It is simple to show that with this expression, the action obtained from
(\ref{local}) remains invariant under the local transformation (\ref{transfm}).
This explicit check
has been discussed earlier in the literature \cite{Mis}.

We now illustrate how the fundamental relation (\ref{law}), which was derived from
general arguments based on the commutativity property, plays a role for
the local symmetries.
Since the generator is given by
(\ref{hamgen}) it might appear that we get a trivial relation $0=0$, which just
follows from the time consistency of the constraints. Such a conclusion is, however,
valid only when we pass to the constraint shell $\phi_a=0$. This is not allowed
in the present context because then the generator itself vanishes. A proper
interpretation of (\ref{law}) is needed meaning that the passage to constraint 
shell is disallowed.
This also entails a slight modification in 
(\ref{neweqn}). The usual hamilton equation following from (\ref{local})
actually contains an extra term which drops out once the constraint 
condition (\ref{constraint}) is imposed. Since we do not want to impose this
condition, the complete equation of motion is,
$$
\dot q_i=\{q_i, \phi_a\}\lambda^a + \{q_i, \lambda^a\}\phi_a +\{q_i, V\}
$$
Putting (\ref{hamgen}) in (\ref{law}) and using the above equation of motion,
along with the constraint algebra (\ref{algebra}), to evaluate $\dot\phi_a$,
we obtain,
$$
\Big(\{\lambda^a, \phi_b\}\e^b-\dot\e^a+C^a_{cb}\lambda^c\e^b+V^a_b\e^b\Big)\phi_a=0
$$
The first term is just $\delta\lambda^a$ following from the definition (\ref{transfm}),
thereby reproducing the condition (\ref{dellambda1}). Both global and local
symmetries may thus be discussed from (\ref{law}). Interpreted this way, it is
possible to regard (\ref{law}) as an analogue of Noether's theorem for the local case.

After completing the hamiltonian analysis we discuss a purely lagrangian approach
which reveals the equivalence of both methods. The first step is to identify the
constraints within the lagrangian formalism. There is a standard method \cite{Bar} of 
doing this thing.
Going back to (\ref{var}) we see that the term in the parentheses must vanish for
the invariance of the action. If there are constraints there will be zero modes of 
the symplectic matrix. Computing these zero modes $\nu^a_\a$ and multiplying from the
left leads to a set of constraints,
\be
\phi^a=(\nu^a_\a)^T\frac{\pa V}{\pa q_\a}=0
\label{zeromodes}\ee
where $T$ stands for the transpose and $a$ is the independent number of zero modes.
These constraints are now inserted in the lagrangian by means of lagrangian multipliers,
so they acquire a form similar to (\ref{local}),
\be
L=a^i(q)\dot q_i+\dot\eta^a\phi_a(q)-V(q)
\label{local1}
\ee
Note however that the constraints have been shifted from the potential to the kinetic part,
implying that $\dot\phi=0$ is being implemented in lieu of $\phi=0$. This ensures the
time consistency of the constraints.
The symplectic matrix with the basic variables $\chi_A=(q_i, \eta_a)$ has the form,
\be
F_{AB}=\left(\begin{array}{cc}f_{ij}&\frac{\pa\phi_a}{\pa q_i}\\
-\Big(\frac{\pa\phi_a}{\pa q_j}\Big)^T&0\end{array}\right)
\label{matrix}\ee
where the first entry is the invertible two form corresponding to the coordinates 
$q_i$ and has exactly the same structure as (\ref{symplectic}).
The zero modes of the above symplectic matrix are given by \cite{{Gov},{Mon}},
\be
\nu^a_A=\left(\begin{array}{cc}f^{ij}\frac{\pa\phi_a}{\pa q^j}\\
\d_{ab}\end{array}\right)
\label{column}
\ee
where $A=i(b)$ for the top(bottom) entry.
Multiplication of (\ref{column}) with $\frac
{\pa V}{\pa q_i}$ to obtain fresh constraints in analogy with (\ref{zeromodes}) just
corresponds to the l.h.s. of (\ref{algebra}). If this turns out to be a combination of
the constraints, the process terminates; else it continues. This is the exact parallel
of the hamiltonian way of extracting the constraints. We assume that the process has
terminated and the lagrangian incorporating all the constraints is given by
(\ref{local1}).

The variation of the lagrangian is given by,
\bear
\d L&=& \Big(f_{ij}\dot q^j+\dot\eta_a\frac{\pa\phi_a}{\pa q^i}-
\frac{\pa V}{\pa q^i}\Big)\d q^i+ \d\dot\eta_a\phi^a \nonumber\\
&=&\Big(F_{AB}\dot\chi^B-\frac{\pa V}{\pa \chi^A}\Big)\d\chi^A
\label{variation1}
\eear 
where $F$ is defined in (\ref{matrix}), and the passage to the second
line from the first has been done by using the commutativity principle.
Let us next discuss the invariance properties.

As emphasised in this approach \cite{Mon} the zero modes generate the infinitesimal transformations,
$$
\d \chi_A=-\e_a(\nu^a_A)^T
$$
which, in components, has the form,
\be
\d q^i=-\e_a\frac{\pa \phi_a}{\pa q^j}f^{ji} ;\,\,\,\d \eta_a=-\e_a
\label{gaugetransfm1}
\ee
Under these transformations the variation (\ref{variation1}) is given by,
\be
\d L= \Big(C^a_{cb}\dot\eta^c-V_b^a\Big)\phi_a \e^b
\label{change}
\ee
which vanishes only if the structure
functions $C$ and $V$ vanish. 
For nonvanishing structure functions, however, the variation of the multipliers
$\eta$
in (\ref{gaugetransfm1}) can be modified such that the r.h.s. of (\ref{change})
vanishes. This is possible since the variation (\ref{change}) is proportional to
the constraints while the first line of (\ref{variation1}) involves a piece
$\d \dot\eta_a\phi_a$. Hence the complete transformation of $\dot\eta$ to achieve
the off-constraint shell invariance of the lagrangian is given by,
$$
\d\dot\eta^a= -\dot\e^a-C^a_{bc}\dot\eta^b\e^c+V_b^a\e^b
$$
Making the necessary identifications $(\lambda=-\dot\eta$) this relation is identical with
(\ref{dellambda1}). This completes the demonstration of the equivalence
between the lagrangian and hamiltonian approaches.

It should be stressed that what has been achieved is the invariance of
the lagrangian (\ref{local}). However this is not the original lagrangian
with which one starts. The latter contains some variables which appear
as lagrange multipliers (say $\lambda_{a_1}$) and implement the constraints
$\phi_{a_1}$ when written in the first order form. A typical example is the
$A_0$ field implementing the Gauss constraint 
in Maxwell's theory. The other multipliers which occur in
(\ref{local}) are put in by hand to enforce the remaining constraints.
Therefore to get the invariance of the original lagrangian, it is essential
to set the remaining multipliers (say $\lambda_{a_2}$) to zero.
 Using (\ref{dellambda1}) this leads to a restriction among the gauge
parameters,
\be
\dot\e^{a_2}=C^{a_2}_{c_1b}\lambda^{c_1}\e^b+V^{a_2}_b\e^b
\label{restriction}
\ee
This equation determines only $a_2$ gauge parameters. Thus the number of
independent free gauge parameters is just $a_1$, namely the number of
original multipliers. The other relation in (\ref{dellambda1}) yields the
variation of the multipliers (the cyclic variables) in the original lagrangian,
\be
\d\lambda^{a_1}=\dot\e^{a_1}-C^{a_1}_{c_1b}\lambda^{c_1}\e^b-V^{a_1}_b\e^b
\label{dellambda3}
\ee
Together with (\ref{transfm}) the above relation yields the symmetry variations on
all the variables in the lagrangian.

It might be mentioned that relations
connecting gauge parameters  were also 
obtained by purely hamiltonian methods \cite{{Henn},{Ban1}} using Dirac's
classification of constraints. However the invariance shown there was
for the total
action which is the original action modified by the inclusion of the
primary constraints; hence those relations  
involved the undetermined multipliers associated 
with the primary constraints. Since these constraints never occur here the undetermined
multipliers also dont occur in our relations.
Also, the invariance shown here is directly with regard to
the original action.
 
We end this section by providing an example with a lagrangian,
\be
L=\frac{1}{2}\Big[\Big(\dot q_2-e^{q_1}\Big)^2+\Big(\dot q_3-e^{q_2}\Big)^2\Big]
\label{example}
\ee
Its first order form is given by,
\be
L=p_2\dot q_2+p_3\dot q_3-p_2e^{q_1}-p_3e^{q_2}-\frac{1}{2}(p_2^2+p_3^2)
\label{firstorder}
\ee
where we use a notation to easily identify the canonical pairs 
following from the symplectic brackets.
The constraints, using either the hamiltonian or the
lagrangian version, are found to be:
$\phi_1=p_2e^{q_1}; \phi_2=p_3e^{(q_1+q_2)}$.
Note that the third term in the r.h.s. of (\ref{firstorder}) is a constraint
and hence it is dropped when we actually implement the constraints through the
lagrange multipliers. Moreover since $q_1$ is a cyclic variable, $e^{q_1}$ is absorbed 
in the multipliers and the final lagrangian
incorporating the constraints is expressed as,
$$
L=p_2\dot q_2+p_3\dot q_3-\eta p_2-\lambda p_3e^{q_2}-p_3e^{q_2}-\frac{1}{2}(p_2^2+p_3^2)
$$
It is simple to check that the modified constraints satisfy the consistency
algorithm.
The variations of the coordinates are given by,
$\d q_2=\e_2;\,\,\, \d q_3=e^{q_2}\e_3$
where $\e_2$ and $\e_3$ are the parameters associated with the two constraints.
To get the invariance of the original action, we have to set $\eta=e^{q_1}$ and
$\lambda=0$. Using (\ref{restriction}) this yields a relation 
$\e_2=\dot\e_3+\e_3\dot q_2$ connecting the two parameters. Also, the variation
of the cyclic variable $q_1$ can be obtained from (\ref{dellambda3}). Using all this
information, the final transformations turn out to be,
$$
\d q_1=e^{-(q_1+q_2)}(\ddot\lambda-\dot\lambda\dot q_2); \d q_2=e^{-q_2}\dot\lambda;
\d q_3=\lambda
$$
where we have redefined $\e_3 e^{q_2}=\lambda$. It is simple to check that (\ref{example})
is invariant under these local transformations.

To conclude, based on the principle of commutativity of a general variation 
with the time differentiation operation, it was possible to discuss global
and local symmetries simultaneously. The Noether result concerning the time
conservation of the generator is therefore applicable
for gauge invariances of either kind. Just as the variational principle plays a 
key role in the lagrangian formulation of symmetries, the most natural way
of understanding the hamiltonian formulation is the commutativity property mentioned
earlier. Since this property is an essential ingredient in deriving the variational
principles, a direct contact between the lagrangian and hamiltonian formulations
was feasible and a complete 
equivalence between the two was established.

\end{document}